\documentclass[a4paper]{jpconf}
\usepackage{graphicx}
\usepackage{hyperref}
\hypersetup{pdfstartview = FitH,
            pdfborder    = {0 0 0.5},
            colorlinks   = false}
\usepackage{color}
\definecolor{green} {rgb} {0, 0.5, 0}  
\newcommand{\rmXA} [2] {\sp{#2} {\rm #1}}
\newcommand{\mchi} {m_{\chi}}
\def \PeriodCa     {19.49  --  79.49}
\def \PeriodCb     {110.74 -- 170.74}
\def \PeriodCc     {201.99 -- 261.99}
\def \PeriodCd     {293.24 -- 353.24}
\def \PlotNumberCa {04949}
\def \PlotNumberCb {14074}
\def \PlotNumberCc {23199}
\def \PlotNumberCd {32324}
\newcommand{\rmF}  {\rmXA{F}  {19}}
\newcommand{\rmXe} {\rmXA{Xe}{129}}
\newcommand{\OnlinePlotNRangHAnnual} [2] {
\href{http://www.tir.tw/phys/hep/dm/amidas-2d/amidas-2d.php%
      ?amidas_2D_function=NR_ang%
      &mode_NR=#1_ang%
      &frame=H%
      &ULab=#2%
      &mode_animation=annual%
      &target=\Target%
      &mchi=\WIMPmass%
      &period=periodC%
      &event_No=500}
     {\fcolorbox {green} {white}
       {\begin{minipage} {15.5 cm}
         \begin{center}
          \vspace{0.1  cm}
           \includegraphics [width = 3.8 cm]
            {#1_ang-\Target-0\WIMPmass-H-0500-\PlotNumberCa-#2}%
           \includegraphics [width = 3.8 cm]
            {#1_ang-\Target-0\WIMPmass-H-0500-\PlotNumberCb-#2}%
           \includegraphics [width = 3.8 cm]
            {#1_ang-\Target-0\WIMPmass-H-0500-\PlotNumberCc-#2}%
           \includegraphics [width = 3.8 cm]
            {#1_ang-\Target-0\WIMPmass-H-0500-\PlotNumberCd-#2}%
          \\ \vspace{0.15 cm}
          #2
          \\ \vspace{0.1  cm}
         \end{center}
        \end{minipage}}}
}
\newcommand{\InsertPlotNRangHAnnual} [2] {
\begin{figure} [t!]
\begin{center}
 \OnlinePlotNRangHAnnual
  {#1}
  {SUPL}
 \\ \vspace{0.6 cm}
 \OnlinePlotNRangHAnnual
  {#1}
  {\LabEAs}
 \\ \vspace{0.2 cm}
 \OnlinePlotNRangHAnnual
  {#1}
  {\LabEu}
 \\ \vspace{0.2 cm}
 \OnlinePlotNRangHAnnual
  {#1}
  {\LabNAm}
 \\ \vspace{0.6 cm}
 \OnlinePlotNRangHAnnual
  {#1}
  {ANDES}
 \\ \vspace{0.2 cm}
 \begin{minipage} {15.5 cm}
  \begin{center}
   {\footnotesize (a) \PeriodCa\ day}\hspace{0.8 cm}%
   {\footnotesize (b) \PeriodCb\ day}\hspace{0.5 cm}%
   {\footnotesize (c) \PeriodCc\ day}\hspace{0.5 cm}%
   {\footnotesize (d) \PeriodCd\ day}
  \end{center}
 \end{minipage}
 \\ \vspace{-0.3 cm}
\end{center}
\caption{
 #2
}
\label{fig:#1_ang-\Target-0\WIMPmass-H-0500-\PlotNumberCa}
\end{figure}
}
\begin{document}
\title{Annual modulations of                            \\ \vspace{-0.15cm}
       the angular recoil--flux/energy distributions of \\ \vspace{-0.15cm}
       WIMP--scattered target nuclei                    \\ \vspace{-0.15cm}
       observed at an underground laboratory}
\author{Chung-Lin Shan}
\address{\it%
         Preparatory Office of
         the Supporting Center for
         Taiwan Independent Researchers                         \\
         P.O.BOX 21 National Yang Ming Chiao Tung University,
         Hsinchu City 30099, Taiwan, R.O.C.\hspace*{-0.1 cm}}
\ead{clshan@tir.tw}
\begin{abstract}
 In this article,
 we compare
 the target and WIMP--mass dependent ``annual modulations'' of
 the angular distributions of
 the recoil flux and energy of
 WIMP--scattered target nuclei
 observed at different underground laboratories.
 For readers' reference,
 simulation plots
 with different WIMP masses
 and frequently used target nuclei
 for all functionable underground laboratories
 can be found and downloaded
 on our online (interactive) demonstration webpage
 ({\tt \url{http://www.tir.tw/phys/hep/dm/amidas-2d/}}).
\end{abstract}
\section{Introduction}

 While most direct Dark Matter (DM) detection experiments
 measure only recoil energies
 of (elastic) WIMP--nucleus scattering events
 deposited in underground laboratory detectors,
 ``directional'' direct detection experiments
 aim to provide 3-dimensional information
 (recoil tracks and/or head--tail senses)
 of WIMP--scattered target nuclei,
 as a promising experimental strategy
 for discriminating WIMP signals from backgrounds
 \cite{Ahlen09, Mayet16, Vahsen21}
 and some incoming--direction--known astronomical events
 \cite{Heikinheimo21}.

 Considering
 the very low theoretically estimated event rate
 and thus
 only a few (tens) of WIMP scattering events
 to be observed per year,
 instead of
 the originally proposed observation of
 the diurnal modulation
 in directional DM detection experiments,
 in Refs.~\cite{DMDDD-N-TAUP2019, DMDDD-P}
 we discussed and compared
 the annual modulations of
 the angular distributions of
 the 3-D velocity (flux) and (average) kinetic energy of
 Galactic halo WIMPs,
 which impinge on our detectors,
 observed at different underground laboratories.
 After that,
 in Refs.~\cite{DMDDD-3D-WIMP-N, DMDDD-NR}
 we finally accomplished
 our double Monte Carlo
 scattering--by--scattering simulation package
 for the 3-D elastic WIMP--nucleus scattering process
 and can provide
 the 3-D recoil direction and then the recoil energy of
 the WIMP--scattered target nuclei
 event by event
 in different celestial coordinate systems.

 Hence,
 in this article,
 we will demonstrate and compare
 (the target and WIMP--mass dependences of)
 the annual modulations of
 the angular distributions of
 the recoil flux
 and the average recoil energy of
 scattered target nuclei
 observed at different underground laboratories.

\section{Annual modulations of
         the angular recoil--flux/energy distributions
         in the horizontal frame}
\label{sec:NR_ang-H}
 \def \LabEAs   {Kamioka}
 \def \LabEu    {LSC}
 \def \LabNAm   {SNOLAB}

 By using our simulation package
 described in detail
 in Ref.~\cite{DMDDD-3D-WIMP-N},
 in Figs.~\ref{fig:NR_ang-F19-0100-H-0500-\PlotNumberCa}
 we show
 the angular distributions of
 the recoil flux of
 $\rmF$ nuclei
 scattered by 100-GeV WIMPs
 in four {\em advanced} seasons
 \cite{DMDDD-N, DMDDD-3D-WIMP-N}
 observed in the horizontal coordinate systems of
 the SUPL,
 the \LabEAs,
 the \LabEu,
 the \LabNAm,
 and the ANDES
 laboratories,
 respectively%
\footnote{
 Interested readers can click each row
 in Figs.~\ref{fig:NR_ang-F19-0100-H-0500-\PlotNumberCa}
 to \ref{fig:NR_ang-Xe129-0200-H-0500-\PlotNumberCa}
 and
 Figs.~\ref{fig:QoN_ang-F19-0100-H-0500-\PlotNumberCa}
 to \ref{fig:QoN_ang-Xe129-0200-H-0500-\PlotNumberCa}
 to open the webpage of
 the animated demonstration
 for the corresponding annual modulation
 (for more considered WIMP masses
  and target nuclei
  as well as
  other underground laboratories).
}.
 5,000 experiments with
 500 total accepted events on average
 (Poisson--distributed)
 in one experiment
 in each 60-day observation period of four advanced seasons
 have been simulated
 and binned into 12 $\times$ 12 bins.

 Moreover,
 in Figs.~\ref{fig:NR_ang-Xe129-0100-H-0500-\PlotNumberCa}
 (and Figs.~\ref{fig:NR_ang-Xe129-0200-H-0500-\PlotNumberCa}),
 we consider $\rmXe$
 as one heavy target nucleus
 (and raise the simulated mass of incident WIMPs
  to $\mchi = 200$ GeV)
 in order to demonstrate
 the target (and the WIMP--mass) dependence(s) of
 the annual modulation of
 the angular recoil--flux distribution.
 Correspondingly,
 in Figs.~\ref{fig:QoN_ang-F19-0100-H-0500-\PlotNumberCa},
 to \ref{fig:QoN_ang-Xe129-0200-H-0500-\PlotNumberCa}
 we show
 the angular distributions of
 the average recoil energy of
 $\rmF$ and $\rmXe$ target nuclei
 scattered by 100- and 200-GeV WIMPs
 in four advanced seasons
 observed in the horizontal coordinate systems of
 five considered laboratories,
 respectively.

 \def \Target   {F19}
 \def \WIMPmass {100}
 \InsertPlotNRangHAnnual
  {NR}
  {The angular distributions of
   the recoil flux of
   $\rmF$ nuclei
   scattered by 100-GeV WIMPs
   in four {\em advanced} seasons
   \cite{DMDDD-N, DMDDD-3D-WIMP-N}
   observed in the horizontal coordinate systems of
   the SUPL,
   the \LabEAs,
   the \LabEu,
   the \LabNAm,
   and the ANDES
   laboratories,
   respectively.%
   \vspace{-0.9 cm}}
 \def \Target {Xe129}
 \InsertPlotNRangHAnnual
  {NR}
  {As in Figs.~\ref{fig:NR_ang-F19-0100-H-0500-\PlotNumberCa},
   except that
   a heavy nucleus $\rmXe$
   has been considered as our target.%
   \vspace{-0.4 cm}}
 \def \WIMPmass {200}
 \InsertPlotNRangHAnnual
  {NR}
  {As in Figs.~\ref{fig:NR_ang-Xe129-0100-H-0500-\PlotNumberCa}:
   $\rmXe$ has been considered as the target nucleus,
   except that
   the mass of incident WIMPs
   has been considered as heavy as $\mchi = 200$ GeV.%
   \vspace{-0.4 cm}}
 \def \Target   {F19}
 \def \WIMPmass {100}
 \InsertPlotNRangHAnnual
  {QoN}
  {The angular distributions of
   the average recoil energy of
   $\rmF$ nuclei
   scattered by 100-GeV WIMPs
   in four advanced seasons
   observed in the horizontal coordinate systems of
   five considered laboratories,
   respectively.%
   \vspace{-1.3 cm}}
 \def \Target {Xe129}
 \InsertPlotNRangHAnnual
  {QoN}
  {As in Figs.~\ref{fig:QoN_ang-F19-0100-H-0500-\PlotNumberCa},
   except that
   a heavy nucleus $\rmXe$
   has been considered as our target.%
   \vspace{-0.4 cm}}
 \def \WIMPmass {200}
 \InsertPlotNRangHAnnual
  {QoN}
  {As in Figs.~\ref{fig:QoN_ang-Xe129-0100-H-0500-\PlotNumberCa}:
   $\rmXe$ has been considered as the target nucleus,
   except that
   the mass of incident WIMPs
   has been considered as heavy as $\mchi = 200$ GeV.%
   \vspace{-0.4 cm}}

 At first,
 in each row (for each laboratory),
 one can observe clearly
 the annual variations of
 the distribution patterns of
 both of
 the nuclear recoil flux
 and the average recoil energy.
 More precisely,
 the all--sky average value of
 the average recoil energy
 (Figs.~\ref{fig:QoN_ang-F19-0100-H-0500-\PlotNumberCa}
  to \ref{fig:QoN_ang-Xe129-0200-H-0500-\PlotNumberCa})
 would be the largest (smallest)
 in the advanced summer (winter)
 for all considered laboratories
 (with all three considered target--WIMP mass combinations).

 Furthermore,
 by comparing
 the distribution patterns of
 both of
 the nuclear recoil flux
 and the average recoil energy
 simulated with different target--WIMP mass combinations
 (for each laboratory),
 one can also find that
 the heavier the target nucleus
 and/or the mass of incident halo WIMPs,
 the wider the distribution patterns,
 since
 the larger the most frequent/energetic recoil angles
 (the angle between
  the nuclear recoil
  and the WIMP incoming velocity),
 caused by
 the stronger the cross section (nuclear form factor) suppression
 \cite{DMDDD-NR}.
 Hence,
 theoretically speaking,
 the 3-D distribution patterns of
 the nuclear recoil flux
 and,
 in particular,
 the average recoil energy
 would be useful
 for constraining (or even identifying) the mass (range) of halo WIMPs.

\begin{figure} [t!]
\begin{center}
 \includegraphics [width = 8.5 cm] {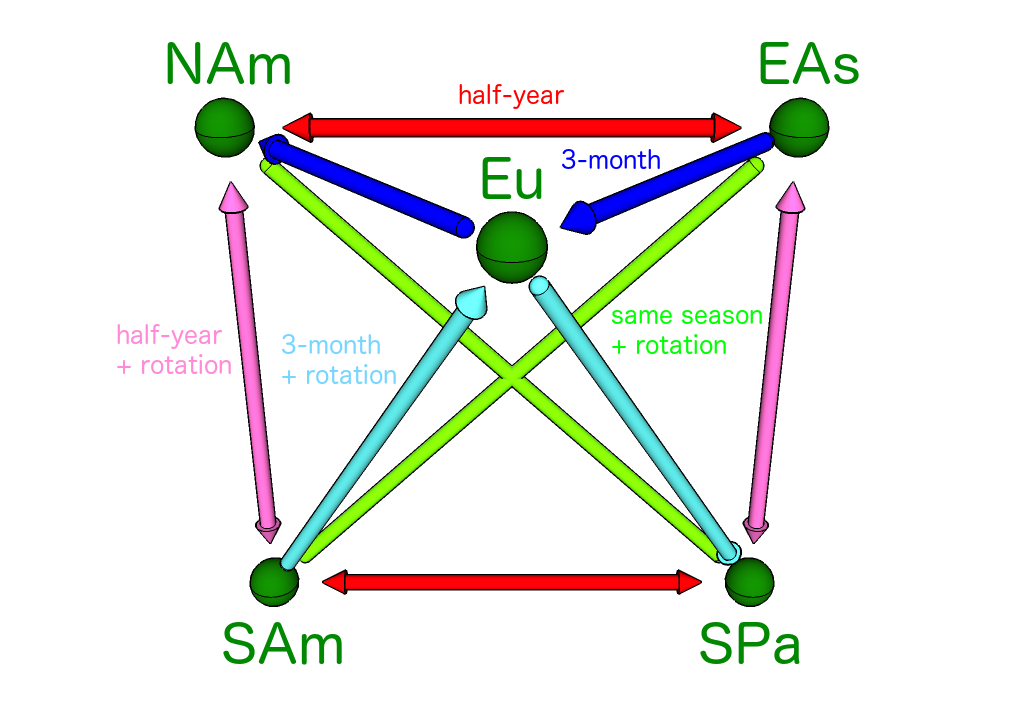} \\ \vspace{-0.5 cm}
\end{center}
\caption{
 The comparisons of
 the angular distribution patterns of
 the 3-D recoil flux and the (average) recoil energy
 observed in the
 horizontal coordinate systems of
 each pair of laboratory groups
 (Figure from Ref.~\cite{DMDDD-P}).
 See the text
 for detailed descriptions.
}
\label{fig:ULabs-Cfs-all-cleared}
\end{figure}

 Finally,
 consider
 the LSC laboratory
 in the European (Eu) group
 as the comparison standard
 (the middle row of
  Figs.~\ref{fig:NR_ang-F19-0100-H-0500-\PlotNumberCa}
  to \ref{fig:QoN_ang-Xe129-0200-H-0500-\PlotNumberCa}),
 as summarized in Fig.~\ref{fig:ULabs-Cfs-all-cleared},
 it can be found that
\begin{itemize}
\item
 the distribution patterns
 observed at Kamioka
 (the East--Asian (EAs) group)
 is similar
 but {\em 3-month earlier};
\item
 the distributions
 observed at SNOLAB
 (the North--America (NAm) group)
 is also similar
 but {\em 3-month later};
\item
 the distributions
 observed at SUPL
 (the South--Pacific (SPa) group)
 is similar to those
 observed at Kamioka
 but with a {\em half--year} difference
 and a 180$^\circ$--degree rotation;
\item
 the distributions
 observed at ANDES
 (the South--America (SAm) group)
 is similar to those
 observed at SNOLAB
 but also with a {\em half--year} difference
 and a 180$^\circ$--degree rotation;
\item
 the distributions
 observed at Kamioka and ANDES
 as well as
 those
 observed at SUPL and SNOLAB
 are similar
 in the {\em same} seasons
 but with a 180$^\circ$--degree rotation.
\end{itemize}
 Note however that,
 while
 the angular distribution patterns of
 the 3-D recoil flux and the (average) recoil energy
 vary from laboratory to laboratory
 (due to their geographical locations),
 the all--sky average values of
 the average recoil energy
 (simulated with each target--WIMP mass combination)
 measured at different laboratories
 in the {\em same} season (observation period)
 should be (almost) equal!

\section{Conclusions}

 In this article,
 we have demonstrated and compared
 the (target and WIMP--mass dependent) annual modulations of
 the angular recoil--flux/energy distributions of
 WIMP--scattered target nuclei
 observed at different underground laboratories.
 Hopefully,
 this work could be useful for our colleagues
 working on
 directional direct Dark Matter detection physics.

\section*{Acknowledgments}
%

 The author would like to thank
 the pleasant atmosphere of
 the W101 Ward and the Cancer Center of
 the Kaohsiung Veterans General Hospital,
 where part of this work was completed.
 This work
 was strongly encouraged by
 the ``{\it Researchers working on
 e.g.~exploring the Universe or landing on the Moon
 should not stay here but go abroad.}'' speech.

\section*{References}
\end{document}